\documentstyle[11pt,psfig]{article}
\newcommand{\ion}[2]{{#1}{\small{#2}}}
\newcommand{\sbb}{mag/$\Box{\rm ''}$}
\def\arcmin{\hbox{$^\prime$}}
\def\arcsec{\hbox{$^{\prime\prime}$}}
\def\farcm{\hbox{$.\mkern-4mu^\prime$}}

\def\la{\mathrel{\mathchoice {\vcenter{\offinterlineskip\halign{\hfil
$\displaystyle##$\hfil\cr<\cr\sim\cr}}}
{\vcenter{\offinterlineskip\halign{\hfil$\textstyle##$\hfil\cr
<\cr\sim\cr}}}
{\vcenter{\offinterlineskip\halign{\hfil$\scriptstyle##$\hfil\cr
<\cr\sim\cr}}}
{\vcenter{\offinterlineskip\halign{\hfil$\scriptscriptstyle##$\hfil\cr
<\cr\sim\cr}}}}}
\def\ga{\mathrel{\mathchoice {\vcenter{\offinterlineskip\halign{\hfil
$\displaystyle##$\hfil\cr>\cr\sim\cr}}}
{\vcenter{\offinterlineskip\halign{\hfil$\textstyle##$\hfil\cr
>\cr\sim\cr}}}
{\vcenter{\offinterlineskip\halign{\hfil$\scriptstyle##$\hfil\cr
>\cr\sim\cr}}}
{\vcenter{\offinterlineskip\halign{\hfil$\scriptscriptstyle##$\hfil\cr
>\cr\sim\cr}}}}}
\title{{\bf Optical \& NIR photometry of the interacting Dwarf Galaxies II\,Zw\,70/II\,Zw\,71
\footnote{Research by P.P, K.J.F. and K.G.N. has been supported by DARA GmbH 
grant 50\ OR\ 9907\ 7 and DFG grant FR 325/50--1. We thank the Calar Alto and La
Palma staff for their assistance during the observations.
}}}
\author{P. Papaderos$^1$, K.G. Noeske$^1$, 
L.M. Cair\'os$^2$, J.M. V\'{\i}lchez$^3$, K.J. Fricke$^1$\\
\vspace{0.1cm}\\
\normalsize $^1$Universit\"{a}ts-Sternwarte G\"{o}ttingen, D-37083 G\"ottingen, Germany\\
\normalsize $^2$IAC, 38200 La Laguna, Tenerife, Spain\\
\normalsize $^3$IAA (CSIC), 18080 Granada, Spain
}
\date{}
\begin{document}
\maketitle
\def\bull{\vrule height .9ex width .8ex depth -.1ex}
\makeatletter
\def\ps@plain{\let\@mkboth\gobbletwo
\def\@oddhead{}\def\@oddfoot{\hfil\tiny
``Dwarf Galaxies and their Environment'';
Bad Honnef, Germany, 23-27 January 2001; Eds.{} K.S. de Boer, R.-J.Dettmar, U. Klein; Shaker Verlag}%
\def\@evenhead{}\let\@evenfoot\@oddfoot}
\makeatother

\begin{abstract}\noindent
We obtained deep optical and NIR images of the pair of
blue compact dwarf (BCD) galaxies II\,Zw\,70 and II\,Zw\,71 
in order to study the effects of interaction on the structural 
properties of their stellar low-surface-brightness (LSB) component. 
We find that within their Holmberg radius the interacting BCDs 
under study do not differ significantly in terms of the central 
surface brightness and exponential scale length
of their LSB hosts from typical iE/nE systems.
In the faint outskirts (26.2$\la\mu_B$\,[\sbb]$\la$28.5) of both systems, however,
the present data reveal conspicuous morphological distortions, most notably 
an extended feature protruding as far as $\sim$9 kpc from the starburst region of 
II Zw 70 in the direction of II Zw 71. The relatively blue colors of this
stellar extension, together with its apparent spatial coincidence
with the massive \ion{H}{I} streamer connecting the dwarf galaxies, are consistent 
with the hypothesis that it originates from recent star formation 
within the gaseous halo of II Zw 70, rather than from stellar matter 
torn out of the LSB host of the BCD during the interaction.
The results presented here support the view that important signatures
of the dynamical response and secular evolution
of the stellar LSB component in interacting dwarf galaxies can be found
in their very faint outskirts.
\end{abstract}
\vspace*{-2mm}
\section{Introduction}
\vspace*{-2mm}
Research on galaxy interactions has so far mainly focussed on 
normal galaxies. However, given the large number ratio of 
dwarf-to-normal galaxies in the local Universe, close 
encounters among gas-rich dwarf galaxies may not be rare
and have a considerable impact on the star formation history 
and dynamical state of these systems. 
Whether interaction/merging with a faint stellar or gaseous companion is the 
genuine process driving starburst activity in blue compact dwarf 
(BCD) galaxies (cf. eg. Taylor et al. 1995, Bergvall et al. 1999) 
is a subject of debate. However, the hypothesis that interactions can 
efficiently trigger star formation activity in gas-rich dwarfs 
received observational support by a number of recent studies 
(e.g. II Zw 33, Walter et al. 1997, M\'endez et al. 1999; 
ESO 338-IG04, \"Ostlin et al. 1998; 
HCG\,31, Iglesias-P\'aramo \& V\'{\i}lchez 1997; Mkn 86, Gil de Paz et al. 2000). 

Here, we present first results from a spectrophotometric study of the 
pair of interacting iI BCDs II\,Zw\,70 and II\,Zw\,71 (Fig. 1a).
At the assumed distance of 18.2 Mpc, the angular separation of 4\farcm 05
between the BCDs corresponds to a projected linear separation of 21.5 kpc.
The ongoing interaction is indicated by interferometric \ion{H}{I} 
\begin{small}
\begin{table}[!t]
\caption{Summary of the observations}
\vspace*{-3mm}
\begin{center}
\begin{tabular}{lcccccc} \hline\hline
Instrument      & \multicolumn{5}{c}{t$_{\rm exp}$\,[min] } \\ 
   &\multicolumn{1}{c}{$U_{\rm J}$}&\multicolumn{1}{c}{$B_{\rm
   J}$}&\multicolumn{1}{c}{$V_{\rm J}$}&\multicolumn{1}{c}{$R_{\rm c}$}&
\multicolumn{1}{c}{$I_{\rm c}$}&\multicolumn{1}{c}{H$\alpha$; H$\alpha$\_cont}\\ \hline
CA2.2/CAFOS     & 60    & 70    & 10    & 30   & 20 & ---\\
NOT/ALFOSC      & 15    & 27    & 22    & 27   & 25 & 25; 25 \\[0.2em]
{\bf Total}     &{\bf 75}&{\bf 97}  &{\bf 32}     &{\bf 57}     &{\bf 45} &{\bf 25; 25}     \\ \hline\hline
                &\multicolumn{3}{c}{II\,Zw\,70}&\multicolumn{3}{c}{II\,Zw\,71}\\ 
                &\multicolumn{1}{c}{$J$}&\multicolumn{1}{c}{$H$}&\multicolumn{1}{c}{$K$\arcmin}&
\multicolumn{1}{c}{$J$}&\multicolumn{1}{c}{$H$}&\multicolumn{1}{c}{$K$\arcmin}\\ \hline
CA3.5/$\Omega$ Prime & 38    & 7     & 17    & 18    & 6     & 22    \\
WHT/INGRID      & 27    & 6     & 8     & 16    & 12    & 8     \\[0.2em]
{\bf Total}   &{\bf 65} &{\bf 13}&{\bf 25}&{\bf 34}&{\bf 18}&{\bf 30} \\ \hline\hline
\vspace*{-1.25cm}
\end{tabular}
\end{center}
\end{table}
\end{small}
studies (Balkowski et al. 1978, Cox et al. 2001) which
revealed a gaseous streamer connecting II Zw 70 with the polar ring galaxy candidate
II Zw 71 (Whitmore et al. 1990, Reshetnikov \& Combes 1994).
To study the effects of interaction on the {\em stellar} component of 
both BCDs we have obtained deep images in 
the optical and NIR (see Table\,1). 
These allow us to investigate the photometric properties of the low-surface-brightness
(LSB) stellar host of both BCDs by means of surface photometry and
search for morphological signatures of the interaction in their outskirts. 
\vspace*{-2mm}
\section{Observations and data reduction}
\vspace*{-3mm}
Optical broad band and H$\alpha$ data of II Zw 70/71
were obtained during several observing runs with the 2.2m Calar Alto 
Telescope equipped with CAFOS and the 2.5m Nordic Optical Telescope (NOT) with ALFOSC.
Near--Infrared exposures were taken with OMEGA PRIME at the
Calar Alto 3.5m telescope, and with the recently commissioned 
INGRID camera at the ING 4.2m William Herschel Telescope (WHT).
All reduced single images in each band were aligned, adapted to equal 
seeing, and coadded weighting each one by its S/N. Optical and NIR
data were calibrated through observations of standards and using field
stars from the 2MASS catalogue.
Surface brightness profiles (Fig. 2ab) were computed 
and decomposed into the luminosity part attributable to the 
underlying (typically exponential) stellar LSB
host and that of the superimposed younger stellar population.
The available data allow for surface photometry studies down to 
a surface brightness level of $\sim$28.5\,\sbb\ and 
24.0\,\sbb\ in $B$ and $J$, respectively.

\begin{figure} 
\begin{picture}(16.5,16.8)
\put(4.8,3){please download 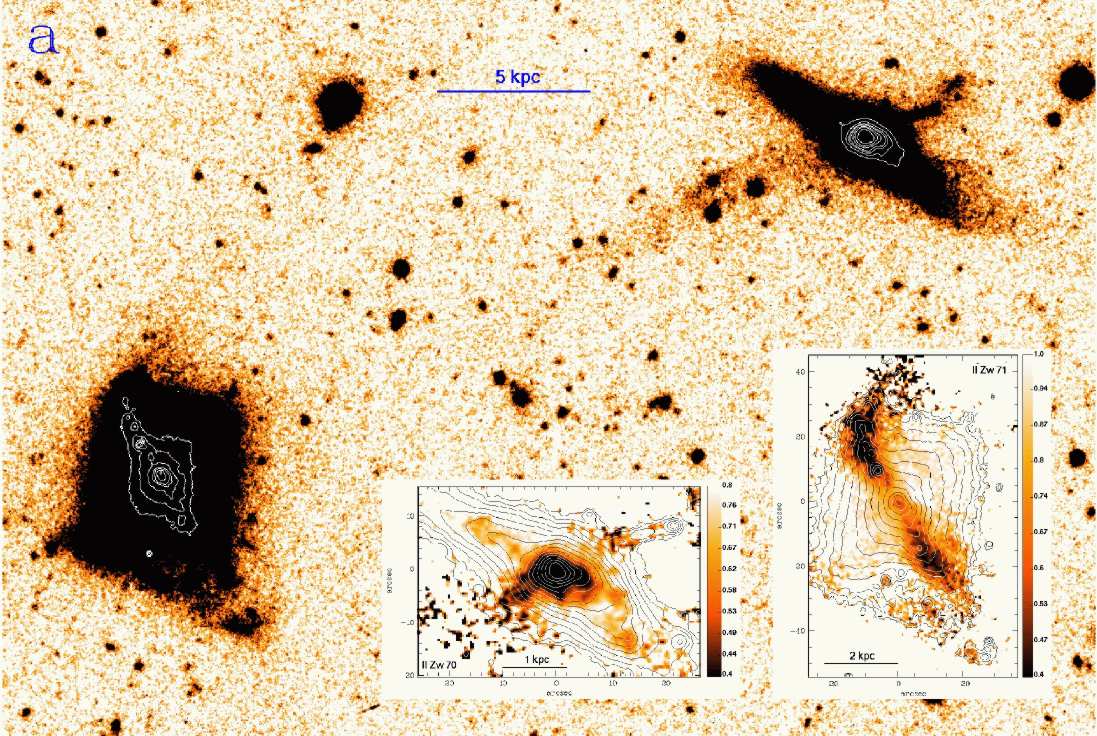 $\dots$ 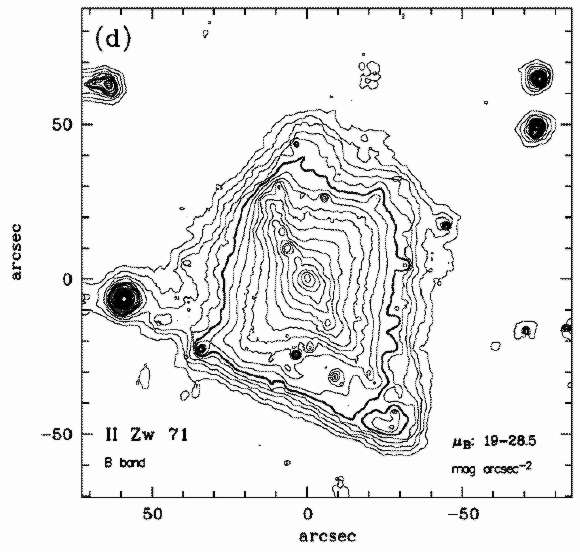}
\put(4.3,2.5){or retrieve the original article from the URL :}
\put(3,2){http://www.astro.uni-bonn.de/$\sim$webgk/dgeproc/dge283.ps.gz}
\end{picture}
\caption[]{{\bf (a)} Combined $B$ band exposure of the BCD pair II Zw
70/71. The field of view is 5\farcm 7$\times$3\farcm 8 and the
total exposure time 1.6 hr. North is up, east to the left. 
The extended asymmetric feature NW\&SE aligned nearly perpendicular 
to the main body of II Zw 70, is visible for surface brightness levels 
$\ga$26 $B$ \sbb. 
The insets show $B-R$ color maps of II Zw 70 and II Zw 71 displayed in 
the range 0.4--0.8 mag and 0.4--1.0 mag, respectively. 
{\bf (b)} \ion{H}{I} map of the pair II Zw 70/71, reproduced from 
Balkowski et al. (1978). A recent interferometric study with the VLA 
(Cox et al. 2001) has shown that the \ion{H}{I} streamer connecting 
II Zw 70 and II Zw 71 has a mass of 2.5$\times$10$^8$ M$_{\odot}$, 
equivalent to 28\% of the total \ion{H}{I} mass of the pair. 
{\bf (c\&d)} Contour maps of II Zw 70 and II Zw 71 
in the $B$ band. In both plots the 26 $B$ \sbb\ isophote 
is indicated by the thick contour.
Note the marked extension labelled SE in the contour map of II Zw 70 
protruding $\ga$ 100\arcsec\ to the direction of II\,Zw\,71.}
\end{figure}
\vspace*{-2mm}
\section{Results and Discussion}
\vspace*{-2mm}
\begin{figure}
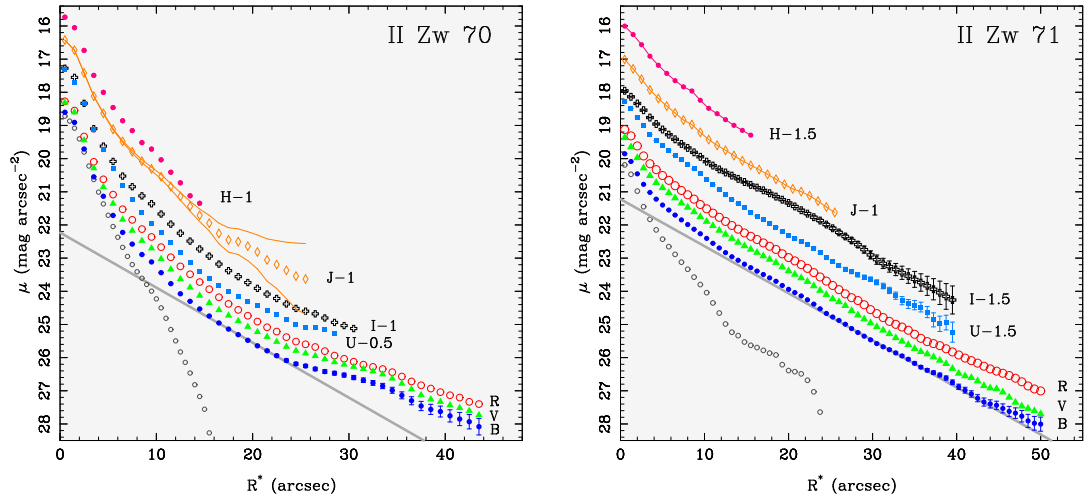
 
\centerline{\psfig{figure=fig2a.ps,height=6.5cm,angle=-90,clip=}
\hspace*{0.5cm}\psfig{figure=fig2b.ps,height=6.5cm,angle=-90,clip=}}
%
\caption[]{Surface brightness profiles from optical and NIR broad band
images. Straight lines show fits to the intensity distribution of the
exponential stellar LSB host of II Zw 70/71 in the $B$ band (see text). 
Small open circles illustrate the surface brightness distribution of the 
emission in excess to the fits and represent the light contribution of the
starburst component. The flattening of the optical profiles of II Zw 70 for
$R^{\star}\ga\,$24\arcsec\ is to be attributed to the 
low surface brightness extensions SE and NW (cf. Fig. 1c).}
\vspace*{-4mm}
\end{figure}
%
The surface brightness profiles of both BCDs can be approximated by an exponential fitting law
at radii where the luminosity contribution of the starburst becomes small
(II Zw 70: 16\arcsec$\la R^*\la$24\arcsec, II Zw 71: 26\arcsec$\la R^*\la$35\arcsec).
For the LSB hosts of II Zw 70 and II Zw 71 we derive, respectively,
a $B$ band central surface brightness of 22.2$\pm$0.15 \sbb\ and 21.2$\pm$0.1 
\sbb\ and an exponential scale length of 580 pc and 675 pc.
Such structural properties are in the range of those inferred previously 
for LSB hosts of iE/nE BCDs (cf. e.g. Papaderos et al. 1996, Cair\'os 2000).
Therefore, no hint towards a significant deviation from the typical 
stellar distribution in the LSB host of BCDs is present, at least on 
a radial scale of $\sim 4$ exponential scale lengths. 
The average colors of the LSB component of either BCD 
(II Zw 70: $B-R$$\la$+0.8 mag, $B-V$=+0.4 mag; II Zw 71:
$B-R$=+1.0 mag, $B-V$=+0.5 mag) are consistent with an evolved 
stellar background with an age of 1--3 Gyr and $\geq 4$ Gyrs for 
II Zw 70 and II Zw 71, respectively (cf. Cair\'os 2000).

At a surface brightness level fainter than $\sim$26.2 $B$ \sbb\ 
the intensity distribution of the less massive interacting 
counterpart, II Zw 70, shows a marked deviation from 
the exponential slope inferred above, getting a scale 
length of $\sim$900 pc and a central surface brightness
of $\sim$23.4 $B$ \sbb. 
This outermost intensity regime, contributing $\sim$18\% of the total 
$B$ band emission of II Zw 70 ($m_B$=14.73 mag), comprises the luminosity 
output of extensions SE and NW (Fig. 1c). 
Roughly 20\% of the emission registered below 26.2 $B$ \sbb\ 
or $\sim$3\% of the total $B$ band light of II Zw 70 is accounted for
by the extension SE. 
This faint feature which can be traced on the available deep
optical exposures out to 1\farcm 65 ($\sim$9 kpc) from the 
optical maximum of II Zw 70 deserves a closer inspection. 
The average colors of the unresolved continuum therein
are substantially bluer ($B-R \sim$+0.4 mag, $V-R \sim +0.16$ mag)
than those determined for the main body of the LSB host.
Such colors lend support to the hypothesis that the SE extension  
is composed of stars having recently formed within the \ion{H}{I} 
streamer connecting the BCDs rather than stars pulled-out from 
the LSB host of II Zw 70 during the interaction. 
The non-detection of H$\alpha$ emission at the location of feature SE (Cair\'os et
al. 2001) in connection with its blue broad band colors are consistent  
with an age of $\sim 10^8$ yr if an instantaneous formation process
is assumed.

The results obtained here support the view that a substantial piece of information 
on the dynamical response and secular evolution of the stellar 
LSB component in interacting dwarf galaxies is potentially present in 
their very faint outskirts.
A continued investigation of the formation process and nature
of the extension SE/NW in II Zw 70/71 as well as the 
search for similar interaction-induced features 
in other pairs of dwarf galaxies 
is apparently of great interest.

{\small
\begin{description}{} \itemsep=0pt \parsep=0pt \parskip=0pt \labelsep=0pt
\item {\bf References}
\item[]Balkowski, C., Chamaraux, P., Weliachew, L. 1978, A\&A 69, 263
\item[] Bergvall, N., \"Ostlin, G., Masegosa, J., Zackrisson, E. 1999, Ap\&SS
  625, 269
\item[]Cair\'os, L.M. 2000, PhD Thesis, Univ. de La Laguna
\item[]Cair\'os, L.M. et al. 2001, ApJS submitted
\item[]Cox, A.L., Sparke, L.S., Watson, A.M., von Moorsel, G. 2001, AJ 121, 692
\item[] Gil de Paz, A., Zamorano, J., Gallego, J. 2000, A\&A 361, 465
\item[] Iglesias-P\'aramo, J., V\'{\i}lchez, J.M. 1997, ApJ 479, 190
\item[] M\'endez, D.I., Cair\'os, L.M., Esteban, C., V\'{\i}lchez, J.M. 1999, AJ
  117, 1688
\item[]\"Ostlin, G., Bergvall, N., Roennback, J. 1998, A\&A 335, 850
\item[]Papaderos, P., Loose, H.--H., Fricke, K.J., Thuan, T.X.  1996, A\&A 314, 59
\item[] Reshetnikov, V.P., Combes, F. 1994, A\&A 291, 57
\item[] Taylor, C.L., Brinks, E., Grashuis, R.M., Skillman, E.D. 1995, ApJS 99, 427
\item[] Walter, F., Brinks, E., Duric, N., Klein, U. 1997, AJ 113, 2031
\samepage\item[] Whitmore, B.C., Lucas, R.A., McElroy, D.B. et al. 1990, AJ 100, 1489
%
\end{description}}
\vspace*{-1cm}
\end{document}